\documentstyle[fleqn,11pt]{article}
\begin{document}
\bigskip
\parindent 1.4cm
\large

\begin{center}

{\Large \bf Dividing Line between Quantum and Classical Trajectories
\vspace{0.1cm}
\\ in a Measurement Problem: Bohmian Time Constant
}
\end{center}
\vspace{1.0cm}
\begin{center}
{Antonio B. Nassar${^1}$ and Salvador Miret-Art\'es${^2}$}
\end{center}
\begin{center}
{\it ${^1}$Science Department, Harvard-Westlake School,
\par 3700 Coldwater Canyon, Studio City, 91604, USA
\par ${^1}$Department of Sciences, University of California,
Los Angeles, Extension Program
\par 10995 Le Conte Avenue, Los Angeles, CA 90024, USA
\par ${^2}$ Instituto de F\'isica Fundamental, Consejo Superior de
Investigaciones Cient\'ificas, Serrano 123, 28006 Madrid, Spain
}
\end{center}
\vspace{1.0cm}
\par
\begin{center}
{\bf Abstract}
\end{center}

This work proposes an answer to a challenge posed by Bell on the
lack of clarity in regards to the line between the quantum and
classical regimes in a measurement problem. To this end, a
generalized logarithmic nonlinear Schr\"odinger equation is
proposed to describe the time evolution of a quantum dissipative
system under continuous measurement. Within the Bohmian
mechanics framework, a solution to this equation reveals a novel
result: it displays a time constant which should represent
the dividing line between the quantum and classical trajectories.
It is shown that continuous measurements and damping not only
disturb the particle but compel the system to converge in time to
a Newtonian regime. While the width of the wave packet may reach a
stationary regime, its quantum trajectories converge exponentially
in time to classical trajectories. In particular, it is shown that
damping tends to suppress further quantum effects on a time scale
shorter than the relaxation time of the system. If the initial
wave packet width is taken to be equal to $2.{\rm{8}} \times
{\rm{1}}{0^{ - {\rm{15}}}} m$ (the approximate size of an
electron), the Bohmian time constant is found to have an upper
limit, i. e., ${\tau _{B\max }} = {10^{ - 26}}s$.

\vspace{1.0cm}
\vspace{0.1cm}
\par E-mail: nassar@ucla.edu
\vspace{1.0cm}

\pagebreak

As pointed out by Bell,\cite{bell} the lack of clarity in regards
to where the transition between the classical and quantum regimes
is located is one aspect of the measurement problem. This problem
represents one of the most important conceptual difficulties in
quantum mechanics. Consequently, this topic of research has gained
considerable interest in the last
decades.\cite{mensky,bialynicki,nassar1} The presence of a
classical apparatus considerably affects the behavior of the
observed quantum system through continuous measurement
\cite{misra,peres,itano,home} which typically fail to have
outcomes of the sort the theory was created to explain. These
frequent measurements are also at the origin of the so-called Zeno
effect.

Another conceptual difficulty is that in a system under
observation there are many degrees of freedom such that
information can be lost in the couplings which may account for
dissipation. One possible approach that has often been used to
answer this question is to introduce all degrees of freedom for
the bath and solve a number of coupled equations in various limits
of some approximation. In fact, by using the influence-functional
method, it has been shown\cite{caldeira} that dissipation tends to
destroy quantum interference in a time scale shorter than the
relaxation time of the system. This result has given justification
for the use of logarithmic nonlinear wave
equations\cite{kostin}-\cite{zhou} to describe quantum
dissipation. These equations have been validated as an
appropriate, practical bath functional in time-dependent density
functional theory for open quantum systems.\cite{zhou}

This work addresses both conceptual difficulties mentioned above.
In particular, an answer to a challenge posed by Bell \cite{bell}
on the dividing line between the quantum and classical regimes in
a measurement problem is given here. To this end, a generalized
logarithmic nonlinear Schr\"odinger equation is proposed to
describe the time evolution of a quantum dissipative system under
continuous measurement. Thus, these two basic existing decoherence
mechanisms are put on equal footing. Nowadays, there are several
routes to deal with this classical-quantum divide. The main three
routes were originally opened up by Bohm \cite{bohm} with his
Bohmian mechanics in 1952, many-worlds interpretation by Everett
in 1957 \cite{everett}, and wave function collapse models
established on firm grounds by Ghirardi, Rimini and Weber in 1986
\cite{grw}. These last authors proposed a unified dynamics (which
has to be stochastic) for microscopic and macroscopic systems,
including the mesoscopic scale. Our approach is much more
restricted by now and follows the first route. Our concerns are
about the time dividing line between classical and quantum
trajectories for quantum processes in presence of different
decoherence mechanisms. The concept of trajectory used in our
context is much more standard. It is limited to a dissipative
(zero temperature) dynamics in presence of continuous
measurements. Furthermore, for the dissipative case, the effective
Hamiltonian we are implicitly considering is of the type of
Caldirola-Kanai Hamiltonian or similar \cite{razavy}. It is worth
noting that we do not propose here a universal behavior. Physical
processes we have in mind are, for example, electronic transport
in materials, diffusion of adsorbates on surfaces or particles in
bulk, motion of particles in quantum viscid media, friction in
qubits, spectral lines under high pressure where the collisions
among gas phase particles can be replaced by a collisional
friction and so on. If, in all cases and any circumstance, an
electron would converge to the classical regime in at most
10$^{-26}$ s, and this time scale would depend on the mass as
$1/\sqrt{m}$ (for negligible friction), one could never, contrary
to a host evidence, observe interference phenomena, with
electrons, neutrons, $C_{60}$, and so on. Such interference
patterns are not considered in our framework of applications. In
such experiments, the measurement is carried out only at the
Fraunhofer or far field region through a screen. Along the way to
the screen, particles are not perturbed by any measurement at any
time.

Within the Bohmian mechanics framework, a solution to this
equation reveals a novel result: it displays a time constant which
establishes the dividing line between the quantum and classical
trajectories. As in RC circuits, the time constant is the key
measure of how quickly the capacitor becomes charged or
discharged; in electronic pacemakers, the pulsing rate of the
heart's contractions is controlled by a RC circuit in which the
time constant represents the most important dividing line between
normal and abnormal heartbeats.\cite{tau} It is shown below that
continuous measurements and damping not only disturb the particle
but compel the system to converge in time to a Newtonian regime
without any assumption of collapse. While the width of the wave
packet may reach a stationary regime, its quantum trajectories
converge exponentially in time to classical trajectories. In
particular, it is shown that damping tends to suppress further
quantum effects on a time scale shorter than the relaxation time
of the system. For example, experiments to measure the size of the
electron consist on colliding two beams of electrons against each
other and counting how many are scattered and altered their
trajectories. By counting the collisions, and knowing how many
particles we have thrown, we can estimate the average size of each
particle in the beam \cite{electron}. If the initial wave packet
width is taken to be equal to $2.{\rm{8}} \times {\rm{1}}{0^{ -
{\rm{15}}}} m$ (the approximate size of an electron), the Bohmian
time constant is found to have an upper limit, i. e., ${\tau
_{B\max }} = {10^{ - 26}}s$.

Bohmian mechanics has recently attracted increasing attention from
researchers.\cite{holland}-\cite{sanz3} Despite the uncertainty
principle, the predictions of nonrelativistic quantum mechanics
permit particles to have precise positions at all times. The
simplest theory demonstrating that this is so is indeed Bohmian
mechanics. One of the fundamental aspects of this mechanics is its
ability to tackle more clearly the quantum measurement problem.
The wave function plays a dual role in this framework; it
determines the probability of the actual location of the particle
and monitors its motion. As pointed out by Bell,\cite{bell} in
physics the only observations we must consider are position
observations - a definite outcome in an individual measurement is
determined by the relevant position variable associated with the
apparatus. It is a great merit of the Bohmian picture to force us
to consider this fact.

For simplicity, let us consider a one-dimensional problem. The
time evolution of the wave function of a quantum dissipative
system $\psi(x,t)$ under continuous measurement can be described
in terms of a nonlinear Schr\"odinger equation. This equation
combines two types of logarithmic nonlinearities: {\it (1)} For
the description of a system under continuous measurement,
Nassar\cite{nassar1} has recently proposed a Schr\"odinger-type
equation with the nonlinear logarithmic term $ - i\hbar \kappa \ln
{\left| {\psi} \right|^2}$, along the lines of the pioneering work
of Mensky\cite{mensky} and Bialynicki-Birula and Mycielski
\cite{bialynicki}, and where the coefficient $\kappa$
characterizes the resolution of the continuous measurement.
However, it is fundamentally different from such an equation due
to the imaginary coefficient in front the logarithmic term. A
remarkable feature of this equation is the existence of exact
soliton--like solutions of Gaussian shape. Hefter \cite{hefter}
has given physical grounds for the use of this logarithmic
nonlinear equation by applying it to nuclear physics and obtaining
qualitative and quantitative positive results. He argues that this
type of equation can be applied to extended objects such as
nucleons and alpha particles. Furthermore, the origin of the
non-linearity can also be understood coming from an energy
dissipation operator in an effective Hamiltonian due to the
continuous measurement or by quantizing the corresponding
Hamilton-Jacobi equation for a linear damped system using the
so--called Schr\"odinger method of quantization
\cite{razavy,sanz3}. {\it (2)} For the description of quantum
dissipative systems, Kostin \cite{kostin} constructed a
Schr\"odinger-type equation with the nonlinear logarithmic term
$(i\nu \hbar /2)\ln (\psi /\psi *)$ with $\nu$ being the friction
coefficient. This equation has the very interesting property that,
at the level of observables, it satisfies the dissipative Langevin
equation at $T = 0$. This equation has subsequently been derived
by Skagerstam\cite{skagerstam} and Yasue\cite{yasue} and has found
extensive use in many applications.\cite{griffin} The Kostin
nonlinear logarithmic term has recently been suggested by
Yuen-Zhou et al. \cite{zhou} as an appropriate, practical bath
functional in time-dependent density functional theory for open
quantum systems with unitary propagation. So, by combining both
nonlinearities, the generalized logarithmic nonlinear
Schr\"odinger equation reads

\begin{eqnarray} \label{eq:schroedinger1}
i\hbar \frac{{\partial \psi (x,t)}}{{\partial t}} = \left[ {H(x,t)
+ i\hbar \left( {{W_c}(x,t) + {W_f}(x,t)} \right)} \right]\psi
(x,t),
\end{eqnarray}
{\bf with}
\begin{eqnarray} \label{eq:schroedinger2}
W_c(x,t) = - \kappa \left[ {\ln {{\left| {\psi (x,t)} \right|}^2}
- \left\langle {\ln {{\left| {\psi (x,t)} \right|}^2}}
\right\rangle } \right]
\end{eqnarray}
and
\begin{eqnarray} \label{eq:schroedinger3}
W_f(x,t) = \frac{\nu }{2}\left[ {\ln \frac{{\psi (x,t)}}{{\psi
*(x,t)}} - \left\langle {\ln \frac{{\psi (x,t)}}{{\psi *(x,t)}}}
\right\rangle } \right],
\end{eqnarray}
The terms in $< >$ arise from the requirement that the integration
of Equation (\ref{eq:schroedinger1}) with respect to the variable
$x$ must be equal to the expectation values of the kinetic and
potential energies through the Hamiltonian $H$ \cite{kostin}. The
expectation value of the energy $<E(t)>$ is defined as in its
standard from
\begin{eqnarray} \label{eq:Aexp}
< E(t) >  \equiv \int\limits_{ - \infty }^{ + \infty } {\psi
^*(x,t)E(t)\psi (x,t)dx}  .
\end{eqnarray}
For the system studied here, no external potential is assumed
(i.e., V = 0).

Equation (\ref{eq:schroedinger1}) has several interesting and
unique properties. First of all, it guarantees the separability of
noninteracting subsystems. Other nonlinear modifications can
introduce interactions between two subsystems even when there are
no real forces acting between them. Second, the stationary states
can always be normalized. For other nonlinearities, stationary
solutions have their norms fully determined and after
multiplication by a constant they cease to satisfy the equation.
And third, the logarithmic nonlinear Schr\"odinger equation
(\ref{eq:schroedinger1}) possesses simple analytic solutions in a
number of dimensions -- especially non-spreading wave-packet
solutions. It is fundamentally different from the equations
proposed by Bialynicki-Birula and Mycielski \cite{bialynicki} due
to $(i)$ the imaginary coefficient in front of the logarithmic
terms and  $(ii)$ the last term $\left\langle {\ln {{\left| {\psi
(x,t)} \right|}^2}} \right\rangle $. Equation
(\ref{eq:schroedinger1}) also generalizes the equation proposed by
Kostin in order to account for continuous observation.

Equation (\ref{eq:schroedinger1}) can now be solved via the
Bohmian formalism.\cite{bohm,holland} To this end, the wave
function is first expressed in polar form:
\begin{eqnarray} \label{eq:wavefunction}
\psi (x,t) = \phi (x,t)\exp (iS(x,t)/\hbar ).
\end{eqnarray}

Now, after substitution of Equation (\ref{eq:wavefunction}) into
Equation (\ref{eq:schroedinger1}), we obtain
\begin{eqnarray} \nonumber
i\hbar \left[ {\frac{{\partial \phi }}{{\partial t}} +
\frac{i}{\hbar }\frac{{\partial S}}{{\partial t}}\phi } \right] =
\end{eqnarray}
\begin{eqnarray} \nonumber
=  - \frac{{{\hbar ^2}}}{{2m}}\left\{ {\left[ {\frac{{{\partial ^2}\phi }}
{{\partial {x^2}}} - \frac{\phi }{{{\hbar ^2}}}
{{\left( {\frac{{\partial S}}{{\partial x}}} \right)}^2}} \right]
+ \frac{i}{\hbar }\left[ {2\frac{{\partial S}}{{\partial x}}
\frac{{\partial \phi }}{{\partial x}} + \frac{{{\partial ^2}S}}
{{\partial {x^2}}}} \right]} \right\}
\end{eqnarray}
\begin{eqnarray} \label{eq:realimag}
- i\hbar \kappa \left[ {\ln {\phi ^2} -  < \ln {\phi ^2} > } \right]
\phi  + \nu \left[ {S - \left\langle S \right\rangle } \right]\phi.
\end{eqnarray}

Equation (\ref{eq:realimag}) can be separated into real and
imaginary parts. By defining the quantum hydrodynamical density
$\rho$, velocity $v$ and quantum potential ${V_{qu}}$ respectively
as
\begin{eqnarray} \label{eq:rho}
\rho (x,t) = \phi^2 {(x,t)},
\end{eqnarray}
\begin{eqnarray} \label{eq:v}
v = \frac{1}{m}\frac{{\partial S}}{{\partial x}},
\end{eqnarray}
\begin{eqnarray} \label{eq:quantumV}
{V_{qu}} =  - \frac{{{\hbar ^2}}}{{2m\phi }}\frac{{{\partial ^2}\phi }}{{\partial {x^2}}},
\end{eqnarray}
we reach

\begin{eqnarray} \label{eq:partialv}
\frac{{\partial v}}{{\partial t}} + v\frac{{\partial v}}{{\partial x}} + \nu v =  - \frac{1}{m}\frac{{\partial {V_{qu}}}}{{\partial x}}
\end{eqnarray}
and
\begin{eqnarray} \label{eq:partialrho}
\frac{{\partial \rho }}{{\partial t}} + \frac{\partial }{{\partial x}}\left( {\rho v} \right) + \kappa \left[ {\ln \rho  - \left\langle {\ln \rho } \right\rangle } \right]\rho  = 0.
\end{eqnarray}

Equation (\ref{eq:partialv}) is an Euler-type equation describing
trajectories of a fluid particle, with momentum $p = mv$, whereas
Equation (\ref{eq:partialrho}) describes the evolution of the
quantum fluid density $\rho$. This density is interpreted as the
probability density of a particle being actually present within a
specific region. Such a particle follows a definite space-time
trajectory that is determined by its wave function through an
equation of motion in accordance with the initial position,
formulated in a way that is consistent with the Schr\"odinger time
evolution. An essential and unique feature of the quantum
potential is that the force arising from it is unlike a mechanical
force of a wave pushing on a particle with a pressure proportional
to the wave intensity. By assuming that the wave packet is
initially centered at $x = 0$ and $\rho (x,0) = {\left[ {2\pi
{\delta ^2}(0)} \right]^{ - 1/2}}\exp \left[ { - {x^2}/2{\delta
^2}(0)} \right]$ and $\rho$ vanishes for $\left| x \right| \to
\infty $ at any time we may rewrite
\begin{eqnarray} \label{eq:rho1}
\rho (x,t) = {\left| {\psi (x,t)} \right|^2} = {\left[ {2\pi
{\delta ^2}(t)} \right]^{ - 1/2}}\exp \left( { - \frac{{{{[x -
\bar x(t)]}^2}}}{{2{\delta ^2}(t)}}} \right),
\end{eqnarray}
where $\delta (t)$ is the total width of the Gaussian wave packet
and $\bar x(t)$ a classical trajectory. Equation (\ref{eq:rho1})
can be readily used to demonstrate that
\begin{eqnarray} \label{eq:average}
\int\limits_{ - \infty }^{ + \infty } {\left( {\left[ {x - \bar x(t)} \right]^2 } \right) {\rho (x,t)} dx}  = \delta ^2 (t).
\end{eqnarray}

Substitution of Equation (\ref{eq:rho1}) into Equation
(\ref{eq:partialrho}) yields
\begin{eqnarray} \label{eq:rhotime}
\frac{{\partial \rho }}{{\partial t}} =
\left[ { - \frac{{\dot \delta }}{\delta } +
\frac{{(x - \bar x)}}{{{\delta ^2}}}\dot {\bar x} +
\frac{1}{{{\delta ^3}}}{{(x - \bar x)}^2} \dot {\delta} } \right]
\rho,
\end{eqnarray}
and
\begin{eqnarray} \label{eq:rhospace}
\frac{{\partial (\rho v)}}{{\partial x}} = \left( {\frac{{\dot \delta }}{\delta } - \kappa } \right)\rho  + \left[ {\left( {\frac{{\dot \delta }}{\delta } - \kappa } \right)(x - \bar x)} + \dot {{\bar x}}\right]\left( { - \frac{{(x - \bar x)}}{{{\delta ^2}}}} \right)\rho,
\end{eqnarray}
which implies that
\begin{eqnarray} \label{eq:v1}
v(x,t) = \left( {\frac{{\dot \delta }}{\delta } - \kappa } \right)(x - \bar x) + \mathop {\bar x}\limits^\cdot.
\end{eqnarray}
Analogously, substitution of Equation (\ref{eq:v1}) into Equation
(\ref{eq:partialv}) yields
\begin{eqnarray} \label{eq:xterms}
\left( {\ddot {\delta} (t) + (\nu - 2\kappa) \dot \delta (t) +
({\kappa ^2} - \kappa \nu) \delta (t) - \frac{{{\hbar
^2}}}{{4{m^2}{\delta ^3}(t)}}} \right){\left( {x - \bar x}
\right)^1} + (\ddot {\bar x} + \nu \mathop {\bar x}\limits^\cdot)
{\left( {x - \bar x} \right)^0} = 0,
\end{eqnarray}
which implies that
\begin{eqnarray} \label{eq:deltaeqn}
{\ddot {\delta} (t) + (\nu - 2\kappa) \dot \delta (t) + ({\kappa
^2} - \kappa \nu) \delta (t) = \frac{{{\hbar ^2}}}{{4{m^2}{\delta
^3}(t)}}}
\end{eqnarray}
and
\begin{eqnarray} \label{eq:quantumxeqn}
\ddot {\bar x} + \nu \mathop {\bar x}\limits^\cdot= 0.
\end{eqnarray}

Equations (\ref{eq:deltaeqn}) and (\ref{eq:quantumxeqn})  show
that continuous measurement of a quantum dissipative wave packet
gives specific features to its evolution: the appearance of
distinct classical and quantum elements, respectively. This
measurement consists of monitoring the position of the quantum
dissipative system and the result is the measured classical
trajectory $\bar x(t)$ for $t$ within a quantum uncertainty
$\delta(t)$.

The associated Bohmian trajectories \cite{wyatt,sanz3} of an
evolving ${i^{th}}$ particle of the ensemble with an initial
position ${x_{oi}}$ can be calculated by first substituting
\begin{eqnarray} \label{eq:xv}
{\dot x_i}(t) = {v_i}(x,t)
\end{eqnarray}
into Equation (\ref{eq:v1}) to obtain
\begin{eqnarray} \label{eq:xtraj1}
{x_i}(t) = \bar x(t) + {x_{oi}}\frac{{\delta (t)}}{{{\delta _o}}}{e^{ - \kappa t}}.
\end{eqnarray}
where $\delta _o=\delta (0)$ is the initial width. As said above,
the position of the center of mass of the wave packet (the
classical trajectory) is represented by ${\bar x(t)}$, while
${x_{oi}}$ is the initial position of the ${i^{th}}$ individual
particle in the Gaussian ensemble corresponding to the wave
function given by Equation (\ref{eq:wavefunction}). Now, Equation
(\ref{eq:deltaeqn}) admits analytic Gaussian-shaped soliton-like
solutions (Gaussons) when
\begin{eqnarray} \label{eq:delta1}
\kappa  = \frac{\nu }{2} + \sqrt {\frac{{{\nu ^2}}}{4} +
\frac{{{\hbar ^2}}}{{4{m^2}\delta _o^4}}}.
\end{eqnarray}
For $\kappa  \ne 0$, and no friction, a stationary regime can be
reached  and the width of the wave packet can be related to the
resolution of measurement which means that if an initially free
wave packet is kept under a certain continuous measurement, its
width may not spread in time. Note that the inverse of
$\frac{\hbar}{{2{m}\delta _o^2}}$ is associated with the relative
spreading of a free Gaussian wave packet \cite{angel}; in other
words, the corresponding spreading velocity is given by
$\frac{\hbar}{{2{m}\delta _o}}$. Thus, the effective or
renormalized friction given by $\kappa$ takes into account the two
spreading mechanisms present in this dynamics. Equation
(\ref{eq:delta1}) displays a very similar structure to that found
for the renormalized frequency of a damped harmonic oscillator. It
is worth stressing that when the friction mechanism is added, the
resolution of the apparatus is changed showing the intertwining
role played by both mechanisms. Finally, the general procedure
will thus be to add more and more independent decoherence
mechanisms in order to take into account the global effect in the
corresponding time evolution of the system, showing again this
entanglement among the different mechanisms.

The transition from quantum to classical trajectories can
then be defined as the Bohmian time constant to be ${\tau _B}
\equiv {\kappa ^{ - 1}}$ and Equation (\ref{eq:xtraj1}) can be
further simplified to
\begin{eqnarray} \label{eq:xtraj2}
{x_i}(t) = \bar x(t) + {x_{oi}}{e^{ - t/{\tau _B}}}.
\end{eqnarray}
It follows from Equation (\ref{eq:xtraj2}) that if ${x_{oi}} = 0$,
then the particle follows the Newtonian trajectory at any time.
If, however, ${x_{oi}}$ is positive, then the particles
distributed in the right half of the initial ensemble are
accelerated whereas the particles distributed in the left half of
the initial ensemble are decelerated. Nevertheless, there is only
a temporary asymmetry in the Bohmian velocities between any two
symmetric particles since the rate of the asymmetry diminishes
with time. After a short time, the distance in position space
shifted by the particles initially lying at positive and negative
${x_{oi}}'s$ converges to a constant value. So, continuous
measurements not only disturb the particle but compel it to
eventually converge to a classical position. It is also noticeable
that damping tends to suppress further quantum effects on a time
scale shorter than the relaxation time of the system. For a small
friction coefficient $(\nu  < \frac{\hbar }{{m\delta _o^2}})$, the
Bohmian time constant can be expressed as
\begin{eqnarray} \label{eq:bohmt}
{\tau _B} \simeq \frac{{2m\delta _o^2}}{\hbar }\left( {1 -
\frac{{\nu m\delta _o^2}}{\hbar }} \right).
\end{eqnarray}

Further, from Equations (\ref{eq:quantumV}) and (\ref{eq:xtraj1})
we have that the quantum force is given by
\begin{eqnarray} \label{eq:force1}
{F_{qu}} =  - \frac{{\partial {V_{qu}}}}{{\partial x}} = -
\frac{\partial }{{\partial x}}\left[ { - \frac{{{\hbar
^2}}}{{8m\delta _o^4}}{{(x - \bar x)}^2} + \frac{{{\hbar
^2}}}{{4m\delta _o^2}}} \right] = \frac{{{\hbar ^2}}}{{4m\delta
_o^4}}{x_{oi}}{e^{ - t/{\tau _B}}}.
\end{eqnarray}
Thus, the convergence of the quantum particle trajectories to
classical trajectories is due to the influence of the measuring
apparatus and friction through the quantum force.\cite{force} This
quantum force is directly proportional to the initial position of
the ${i^{th}}$ particle and decays exponentially in time (it drops
$63\%$ of its initial value after a time constant ${\tau _B}$).
Likewise, the quantum position $x_i(t)$ - the initial position of
the ${i^{th}}$ individual particle in the Gaussian ensemble -
approaches its classical value. So, friction and continuous
observation of a wave packet may lead to a gradual freezing of the
quantum features of the particle.

Finally, if the initial wave packet width for an electron is taken
to be equal to $2.{\rm{8}} \times {\rm{1}}{0^{ - {\rm{15}}}} m$
(the approximate size of an electron \cite{electron}) and the
coefficient of friction is made very small $(\nu  <  < \frac{\hbar
}{{m\delta _o^2}})$, the Bohmian time constant is found to have an
upper limit:
\begin{eqnarray} \label{eq:taumax}
{\tau _{B\max }} = {10^{ - 26}}s.
\end{eqnarray}
This result provides an answer to a challenge posed by Bell
\cite{bell,bernstein} on the lack of clarity about the line
between the quantum and classical regimes in a measurement
problem: The Bohmian time constant above defined may establish
that dividing line.\cite{tau}

By adopting the Bohmian framework, which is one of the three main
routes mentioned at the beginning, the interpretational scheme is
different but pointing in the same direction as other works on
decoherence. Furthermore, we also surmise that that there is no
single universal time scale, but several ones depending on the
experimental situation. On the other hand, the modeling of
effective collapse induced by non-linearity at the quantum level
is scarcely in this route. In any case, further investigation is
needed in order to better understand the dynamics of a system
interacting with an environment, which is traced out, by
considering stochasticity through additional noise terms.
 \pagebreak

{\bf Acknowledgments}

Part of this work was done during the Summer 2012 at the UCLA
Physics and Astronomy Department. Correspondence with I.
Bialynicki-Birula is gratefully acknowledged. S. M-A grateful
acknowledges the MICINN (Spain) through Grant FIS2011-29596-C02-01
and the COST Action MP1006 "Fundamental Problems in Quantum
Physics". Finally we would like to thank two anonymous referees
for their pertinent remarks and suggestions.


\begin{thebibliography}{99}
\bibitem{tau} One of the most interesting uses of an RC-circuit is the
electronic pacemaker, which can make a stopped heart start beating
again. The electrodes are implanted in the heart and the circuit
contains a capacitor and a resistor. The pulsing rate rate depends
on the time constant $\tau = RC$ which regulates the frequency of
the heartbeat.
\bibitem{force} From Equation (\ref{eq:rho1}), we note that the
expectation value of the quantum force vanishes at all times:
$\left\langle {{F_{qu}}} \right\rangle  =  - \frac{1}{m}\left\langle
{\frac{{\partial {V_{qu}}}}{{\partial x}}} \right\rangle  =
\frac{{{\hbar ^2}}}{{2m{\delta ^2}}}\left\langle {x - \bar x(t)}
\right\rangle  = 0$.
\end{thebibliography}
\end{document}